\documentclass{ws-ijmpa}

\begin{document}

\markboth{P.D.Stack and R.Delbourgo}
{GR of YM}

\title{The General Relativity of Two Properties}

\author{Paul D Stack and Robert Delbourgo}

\address{School of Mathematics and Physics, University of Tasmania, 
Private Bag 37 GPO\\ Hobart, Tasmania 7001,AUSTRALIA 7001\\
pdstack@utas.edu.au, bob.delbourgo@utas.edu.au}

\date{\today}% It is always \today, today,

\maketitle

\begin{abstract}
We demonstrate that the extension of the space-time  metric to incorporate two 
anti-commuring property coordinates automatically leads to the unification 
of gravity with nonabelian gauge theory, as well as producing a cosmological term.
\end{abstract}

\section{\label{sec1} Introduction}

Space-time describes a `when' and a `where' of an event but not a `what'. 
The traditional approach for addressing the character of an event is to 
introduce various quantum fields with associated quantum numbers and 
enforce conservation laws on the combination of quantum numbers. While this 
approach most certainly works, it leaves us with many open questions,
such as why multiple generations of particles exist or why they have their 
respective masses. Over a number of years we have attempted to introduce
 some mathematics into the {\em attributes} of an event by attaching property 
 coordinates for characterizing the `what'. These property coordinates are to be 
 associated with Lorentz scalar anti-commuting quantities carrying quantum 
 numbers; the conservation laws for events arise simply by Grassmann 
 integration over the property coordinates in much the same way that overall 
 momentum conservation comes about by integrating over space-time in 
local field theory. The property coordinates provide a mechanism for explaining 
why there are repeated generations of particles and they potentially 
answer\cite{RDPJRW,RDRZ,RD2,RD3,RD4} other questions left open in the 
Standard Model.

In a previous paper\cite{empaper}, we discussed and formalized the general 
extension of space-time which comes about by attaching multiple property 
coordinates; this led us to General Relativity (GR) on a graded manifold.
We later confined ourselves to  a single property coordinate and constructed 
a supermetric that transformed correctly under local gauge transformations. 
Using this supermetric we determined the Ricci supertensor and Ricci superscalar 
and discovered that the Einstein Field equations for gravity plus electromagnetism 
ensued as well as a cosmological constant. This paper will be an 
extension of that work, looking at the non-abelian case with two property coordinates. 
and we shall see that this leads to a unification of gravity with Yang-Mills theory 
whereby the stress-tensor for the non-abelian gauge fields arises as part of the 
generalized Ricci tensor.

\section{Property coordinates, notation and supermetric}

Following our earlier work we will attach to space-time two complex anti-commuting 
coordinates $\zeta$ and their conjugates $\bar\zeta$. We choose these coordinates 
to be Lorentz scalar, which distinguishes our scheme from supersymmetry (where
 the extra degrees of freedom transform as spinors). The coordinates are anti-commuting 
 so we also avoid infinite particle states, because series expansions in the coordinates 
 will terminate. Quantum numbers are attributed to these property coordinates and 
 particle fields are produced via superfield expansions. Consequences of such a 
 scheme with five property coordinates are covered by papers\cite{RD2,RD3,RD4}. 
In this paper we will be focussed purely on the geometrical aspect, namely what happens
to general relativity when two of these property coordinates are included.

As before, we adopt the convention that the extended space-time-property indices run over 
uppercase Roman characters $(M,N,L,$ etc), space-time indices run over lowercase 
Roman characters $(m,n,l,$ etc) and property indices run over Greek letters 
$(\mu,\nu,\lambda$, etc). The grading of an index is given by $[M]$, where 
$[m]=0$ for even indices and $[\mu]=1$ for odd indices.

We intend to model our notation on standard GR as far as possible, but
this does produce some tension with particle physics notation. Consider 
a property coordinate $\zeta^\mu$; the conjugate of this coordinate in usual particle 
physics parlance would be written $\bar\zeta_\mu$. On the other hand, raising and lowering
indices in GR  corresponds to swapping between contravariant and covariant coordinates. 
Since we are introducing Lorentz {\em scalar} property coordinates  and want
to maintain GR notation, we we will adopt the following algebraic convention: 
$\zeta^\mu = \zeta_{\bar\mu}$ and $\zeta^{\bar\mu} = -\zeta_{\mu}$. The minus sign 
exists to ensure self-consistency of the up then down summation of our graded GR with 
the bar and no bar summation of particle physics; it will also guarantee that we sum 
barred indices with unbarred ones when they both occur as superscripts or subscripts, in
order to produce internal symmetry invariants.

To construct our metric we first consider a flat manifold possessing coordinates 
${X^M} = (x^m,\zeta^\mu,\zeta^{\bar\mu})$ without space-time curvature or gauge fields. 
Our metric distance in this case is given by:
\[ ds^2 = dX^M dX^N\eta_{NM} = dx^m dx^n\eta_{nm} +  \ell^2 (d\zeta^\mu 
d\zeta^{\bar\nu} \eta_{\bar\nu \mu}+d\zeta^{\bar\mu} d\zeta^\nu \eta_{\nu\bar\mu})/2
 \]
where $\eta_{\mu \bar\nu} = -\eta_{\bar\nu \mu} = \delta_\mu{}^\nu $ and $\eta_{nm}$ is 
Minkowskian. The property coordinates $\zeta$ and $\bar\zeta$ are being taken as 
dimensionless, so a length scale $\ell$ has been introduced to ensure the metric distance 
carries the correct units. If we make a local non-abelian unitary property coordinate 
transformation of the form:
\begin{equation} \label{propertyphasetrans}
x^\prime = x;  \;\;\;
 \zeta^{\mu\prime} = [e^{i \Theta(x)}]^{\mu\bar\nu} \zeta^\nu; \;\;\;
 \zeta^{\bar\mu\prime} = \zeta^{\bar\nu} [e^{-i \Theta(x)}]^{\nu\bar\mu} \;\;,
\end{equation}
the above metric does {\em not} transform correctly as a tensor strictly should. To fix this 
problem it is essential to introduce non-abelian gauge fields $W_m{}^{\mu\bar\nu}$ 
(accompanied by a coupling constant $e$), through frame vectors as described in  
Appendix A.1.  This yields the following metric:
\begin{equation}
G_{MN} = \left(
\begin{array}{ccc}
g{}_{m}{}_{n}+\frac{1}{2} e{}^2  l^2  \bar\zeta (W_{m}W_{n}+W_{n}W_{m}) \zeta  & 
-\frac{1}{2} i e l^2 (\bar\zeta W_{m}){}^{\bar{\nu}} & -\frac{1}{2} i e l^2 (W_{m}\zeta){}^{\nu} \\ 
-\frac{1}{2} i e l^2 (\bar\zeta W_{n}){}^{\bar{\mu}}& 0 & \frac{1}{2} l^2 \delta{}^{\nu}{}^{\bar\mu} \\ 
-\frac{1}{2} i e l^2 (W_{n}\zeta){}^{\mu} & -\frac{1}{2} l^2 \delta{}^{\mu}{}^{\bar\nu} & 0.
\end{array}
\right)
\end{equation}
The reader may readily check, from the transformation properties of the supermetric under 
coordinate changes, that gauge covariance now follows naturally.
In fact (2) is not the most general metric that transforms correctly under a local property phase
transformation. Without including (ghost) fields that anticommute, which are likely to
arise in a quantised version, the metric can be generalised by
allowing for products of the property coordinates that are invariant under local phase 
transformations; these act like property curvature terms: 
$\zeta^{\bar\mu} \zeta^\mu$. 
This results in the following metric, with four independent curvature coefficients $c_i$:
\begin{equation} \label{metricform}
G_{MN} = \left(
\begin{array}{ccc}
G_{m n} & G_{m \nu} & G_{m \bar\nu}  \\ 
G_{\mu n} & 0 & G_{\mu \bar\nu} \\ 
G_{\bar\mu n} & G_{\bar\mu \nu} & 0
\end{array} \right)
\end{equation}
where
\begin{align}
G_{mn} =& g{}_{m}{}_{n}\big(1+c_1 \bar\zeta\zeta + c_2 (\bar\zeta\zeta)^2\big) 
+\frac{1}{2} e{}^2  l^2  \bar\zeta (W_{m}W_{n}+W_{n}W_{m}) \zeta \big(1+c_3 \bar\zeta \zeta\big) \nonumber\\
G_{m\nu} =& -\frac{1}{2} i e l^2 (\bar\zeta W_{m}){}^{\bar{\nu}} \big( 1 + c_3 \bar\zeta \zeta\big) \nonumber\\
G_{m\bar\nu} =&  -\frac{1}{2} i e l^2 (W_{m}\zeta){}^{\nu} \big( 1 + c_3 \bar\zeta \zeta\big) \nonumber \\
G_{\mu\bar\nu} =&  \frac{1}{2} l^2 \delta{}^{\nu}{}^{\bar\mu} \big(1+c_3 \bar\zeta\zeta 
+ c_4 (\bar\zeta\zeta)^2\big). \label{2coordmetric}
\end{align}
The requirement that the metric transforms correctly as a tensor under local
 property phase transformations has severely restricted the nature of the metric. 
 As can be seen from (\ref{2coordmetric}), there are only four free parameters 
 introduced by incorporating as many U(2) invariants as possible into the metric\footnote{
 If we want to separate the U(1) and SU(2) couplings, as may be required for electroweak theory,
 then we must replace $eW\zeta(1+c\bar{\zeta}\zeta)$ by $e'W'I\zeta(1+c'\bar{\zeta}\zeta) +
 e\underline{W}.\underline{\sigma}\zeta(1+c\bar{\zeta}\zeta)$. This complication is left for future work.}. 
 We envisage that the $c_i$'s are the expectation values of chargeless 
 Higgs or dilaton fields. [Fermionic fields would carry a Lorentz spinor index  which 
 would conflict with Lorentz invariance if they made an appearance in the $G_{m\nu}$ sector.]

The subsequent formulae are simplified somewhat by scaling all the curvature 
coefficients to $c_3\equiv c$ as follows,
$c_1=b_1c,\,\, c_2=b_2c^2, \,\, c_4=b_4c^2,$
and by defining $b \equiv 2b_1-b_2+2b_4-3$. In particular the superdeterminant of 
this metric can be found by considering eqn. 36 from our previous 
paper\cite{empaper} and reduced to
\begin{equation} \label{sdetformula}
s\det(X) = \det(A-B D^{-1} C) \det(D)^{-1}.
\end{equation}
Applying this to our metric results in the particular form,
\begin{eqnarray}
\sqrt{-G_{..}}& =&  \frac{4}{l^4} \sqrt{-g} \big[ 1 + 2(c_1 - c_3) \bar\zeta \zeta+ 
(c_1{}^2 + 2c_2 - 4 c_1 c_3 + 3 c_3{}^2 - 2 c_4)  (\bar\zeta \zeta)^2 \big]\nonumber\\
& = & \frac{4}{l^4} \sqrt{-g}\big[1+2(b_1-1)c\bar{\zeta}\zeta+
(b_1{}^2+2b_2-4b_1+3-2b_4)(c\bar{\zeta}\zeta)^2\big].
\end{eqnarray}

\section{Ricci supertensor and superscalar curvature}
Using eqn. 14 from our previous paper\cite{empaper}, namely:
\begin{equation}\label{gammaformula}
 \Gamma_{MN}{}^K \!\!=\!\!\big[ (-1)^{[M][N]\!+\![L]} G_{ML,N}
  + (-1)^{[L]} G_{NL,M}- (-1)^{[L][M]\!+\![L][N]\!+\![L]} G_{MN,L} \big] G^{LK}/2,
\end{equation}
and the metric tensor (4), we can evaluate the 
Christoffel symbols $\Gamma_{MN}{}^{K}$. These are listed in the Appendix A.3.
From the Christoffel symbols the Ricci supertensor ${\cal R}_{KM}$ can be found 
using eqn. 18 from our previous paper\cite{empaper}:
 \begin{eqnarray}
{\cal R}^J{}_{KLM} &=& (-1)^{[J]([K]+[L]+[M])}  \big[(-1)^{[K][L]} 
\Gamma{}_{K M}{}^J{}_{,L} - (-1)^{[K][M]+[L][M]} \Gamma{}_{K L}{}^J{}_{,M}
\nonumber \\
&&\qquad\qquad\qquad\qquad + (-1)^{[L][M]} \Gamma{}_{KM}{}^R \Gamma{}_{R L}{}^J
-\Gamma{}_{KL}{}^R\Gamma{}_{RM}{}^J \big], \label{riemannformula}
\end{eqnarray}
and then taking the trace to get 
\begin{equation}
{\cal R}{}_{KM} = (-1)^{[K][L]}{\cal R}^L{}_{KLM}.
\end{equation}
Finally the Ricci superscalar ${\cal R}$ can be found by contracting with the metric to get
\begin{equation}\label{rdef}
{\cal R} = G^{MK} {\cal R}_{KM}.
\end{equation}

We are now in a position to evaluate the total Lagrangian density,
\begin{equation}
{\cal L}=\int\!\! d^2\zeta d^2\bar\zeta \,\sqrt{-G..}{\cal R} \,\,.
\end{equation}
The evaluation is aided by an algebraic computer program coded in Mathematica. This results in
\begin{eqnarray}
{\cal L} &=& \sqrt{-g..}  \left[ \frac{8c^2 b}{l^4} R^{[g]} -
\frac{e^2c}{l^2} {\rm Tr}\big(\mathcal{F}^{m n} \mathcal{F}_{mn}\big)\right. \nonumber\\
& & \left.+\frac{16c^3}{l^6}
(-24b_1b_2+38{b_1}^2+40b_2 -110b_1+75 +40 b_1b_4- 60b_4)\right], \label{lagdens}
\end{eqnarray}
where $R^{[g]}$ is the gravitational space-time Ricci scalar and 
$\mathcal{F}_{mn}$ is the non-abelian field tensor,
 $\mathcal{F}_{mn}= W_{n,m} - W_{m,n} -  i e [W_m, W_n]$.
We can now recognize 
\begin{align}
&16\pi G_N\equiv  \frac{e^2 l^2}{ 2bc}, \qquad b=2b_1-b_2+2b_4-3;\\
& \Lambda = c(24b_1b_2-38 b_1{}^2-40b_2+110 b_1-75-40 b_1b_4+ 60b_4)/ l^2 b.
\end{align}

We see that the Yang-Mills Lagrangian arises naturally from the geometry, 
as well as a cosmological constant $\Lambda$. In our previous paper with 
one property coordinate we only had two free curvature parameters in the metric, which 
produced a negative cosmological constant. Now with two property coordinates and
the possibility of $(\bar{\zeta}\zeta)^2$ terms we 
have the opposite problem: there are four free curvature parameters and 
our cosmological constant is essentially unrestricted, though it is at least now 
consistent with a positive value\footnote{Observe that some simplification takes place 
if one chooses $b_2={b_1}^2$ and $b_4=b_1$ but there is really no compelling reason
for that choice at this stage.}. We envisage there will be additional symmetry 
considerations, when we examine reparametrizations of $\zeta$ or which arise from 
quantisation, that will restrict the possible values for the cosmological constant.

We can now look at variation of the Lagrangian density with respect to 
the space-time metric and the gauge field to produce field equations; this will 
provide a check that everything is working consistently at the level of
the extended superspace $X$ and with the Mathematica coding. The variation of 
the supermetric with respect to the space-time metric is simply just 
$\delta G_{mn}= (1+c_1\bar\zeta\zeta+c_2 (\bar\zeta\zeta)^2\big)\delta g_{mn}$. 
The variation of the Lagrangian density with respect to the space-time metric is 
then given by
\begin{align}
&\int d^2\zeta d^2\bar{\zeta} 
\sqrt{-G_{..}} (R^{mn}-\frac{1}{2} G^{mn}R) \delta G_{mn}/\delta g_{mn}=\nonumber\\
&\frac{16c^3}{l^6} (12 b_1 b_2-19 b_1{}^2-20 b_2 +55 b_1
-75/2-20 b_1 b_4+30 b_4) g^{mn}\nonumber\\
&+\frac{8c^2b}{ l^4}\left(R^{[g]\;mn} - \frac{1}{2} g^{mn} R^{[g]}\right)
- \frac{2ce^2}{l^2}{\rm Tr}(\mathcal{T}^{mn}),
\end{align}
where $\mathcal{T}^{mn} = \mathcal{F}^{m l} \mathcal{F}^{n}{}_{l}
-\frac{1}{4} g{}^{m}{}^{n}\mathcal{F}^{l k} \mathcal{F}_{l k}$ is the 
Yang-Mills stress tensor. Equating this to zero results in the following field equations:
\begin{align}
&R^{[g]\;mn} - \frac{1}{2} g^{mn} R^{[g]}- \frac{e^2l^2}{4bc} {\rm Tr}(\mathcal{T}^{mn})\nonumber\\
&+\frac{c}{l^2b}(24b_1b_2-38b_1{}^2-40b_2+110b_1-75-40b_1b_4+60b_4)g^{mn}=0,
\end{align}
which is consistent with the Lagrangian in (12).

The next verification is the variation of the Lagrangian density with 
respect to the gauge field. First we express our gauge field in terms of a basis, 
$W_m{}^{\bar\mu \nu} = W_{m }{}^i \tau_i{}^{\bar\mu\nu}$, where 
$\underline\tau = (I,\underline\sigma)$. We then consider the variation 
of our metric $G_{MN}$ with respect to $W{}_{p}{}^{i}$. 
\begin{align}
\delta G_{mn} =&\frac{1}{2} e^2  l^2  \bar\zeta \left(
\delta_{m}{}^{p} \tau_i W_{n}
+\delta_{n}{}^{p} W_{m}\tau_i  
+\delta_{n}{}^{p} \tau_i W_{m}
+\delta_{m}{}^{p}W_{n}\tau_i \right) \zeta \big(1+c\bar\zeta \zeta\big) 
\;\delta W_{p}{}^{i} ,\nonumber\\
\delta G_{m\nu} =& -\frac{1}{2} e^2 (\bar\zeta \tau_i){}^{\bar{\nu}} 
\big(1+c \bar\zeta \zeta\big)\delta_{m}{}^p \delta W_{p}{}^i ,\nonumber\\
\delta G_{m\bar\nu} =&  -\frac{1}{2}e^2 (\tau_i \zeta){}^{\nu} 
\big(1+c\bar\zeta \zeta\big)\delta_m{}^p \delta W_{p}{}^i  .\nonumber
\end{align}

Next, we find that the variation of the Lagrangian density with respect to $W_p{}^{i}$
does not involve curvatures coefficients other than $c_3=c$:
\begin{align}
&\int d^2\zeta d^2\bar\zeta \sqrt{-G_{..}} (R^{MN} - 
\frac{1}{2} G^{MN}R) \delta G_{NM}/ \delta W^{n} \nonumber\\
&\propto (W_{m,n}-W_{n,m})^{,m}+2ie[W^m,W_{n,m}] +ie[W_{m,n},W^m]\nonumber\\
&\,\,\,+ie[{W^m}_{,m},W_n]
+e^2(W^mW_mW_n-2W^mW_nW_m+W_nW^mW_m)
\end{align}
Equating this to zero  gives precisely the non abelian version of the Maxwell 
equations in free space, namely ${\cal D}^m {\cal F}_{mn} =0$. This reassures
us that the work is free of errors. Clearly adding curvature to space-time as well
just gives the general relativistic versions of the equations of motion.

\section{Source Superfields}
To close the circle we need to ascertain that in our formalism scalar and spinor 
sources lead to the correct non-abelian interactions of the gauge fields with those sources.
To that end it suffices to ignore all curvature constants except for $c_3=c$, which
enters the space-time property sector of the metric and accompanies $eW$. In
this context we might regard attribute 1 as referring to neutrinicity and attribute 2
as referring to electricity to get an idea of the properties of the resulting
expansions in property.

\subsection{Scalars}
To begin with ignore all gauging (and thus all curvatures in $G$). In that flat limit an 
anti-selfdual\cite{RD4} scalar superfield admits an expansion that produces a singlet $Y$ and a 
triplet $\underline{Z}$ of SU(2):
\begin{eqnarray}
\sqrt{2}\Phi &=& Y[1-(\bar{\zeta}\zeta)^2/2]+Z^0(\zeta^{\bar{1}}\zeta^1-\zeta^{\bar{2}}\zeta^2)
    + Z^+\zeta^{\bar{1}}\zeta^2 + Z^-\zeta^{\bar{2}}\zeta^1\nonumber \\
    & = & Y[1-(\bar{\zeta}\zeta)^2/2] + \bar{\zeta}\underline{Z}.\underline{\sigma}\zeta.
\end{eqnarray}
Using the lemma $\int d^2\zeta d^2\bar{\zeta} \,\,(\bar{\zeta}A\zeta)(\bar{\zeta}B\zeta)
= {\rm Tr}(A){\rm Tr}(B)-{\rm Tr}(AB)$, we readily establish the normalization,
\begin{equation}
\int d^2\zeta d^2\bar{\zeta}\,\, \Phi^2(X) = -(Y^2 + \underline{Z}.\underline{Z}).
\end{equation}
Next we introduce curvature into the metric, but limited just to checking that the
gauge interactions emerge correctly. Thus we discard all $c_i$ except for $c_3=c$,
which multiplies $eW$; in that case the Berezinian collapses to 
\[
 \sqrt{G..} \rightarrow 4[1-2c\bar{\zeta}\zeta+3(c\bar{\zeta}\zeta)^2]/l^2
\]
and the kinetic Lagrangian reduces to the property integral,
\begin{eqnarray}
\int d^2\zeta d^2\bar{\zeta}\sqrt{G..}\,G^{MN}\partial_N \Phi\partial_M \Phi &=&
-(1-3c^2)\partial^m Y\partial_m Y\nonumber \\
&&- {\rm Tr}\big[\!(\partial^mZ\!-\!ie[W^m\!,Z]).(\partial_mZ\!-\!ie[W_m,Z])\!\big].
\end{eqnarray}
This is exactly what we would have expected for the interaction of the four scalar fields
with the gauge field apart from a trivial renormalization of the $Y$-field.

\subsection{Spinors}
So far, so straightforward. However the fermion sources prove more difficult to handle.
This is because the vielbein generalization for the `Dirac' operator, viz. 
$\Gamma^A {E_A}^M\partial_M$ introduces a new set of `gamma-matrices'
$\Gamma^\alpha$, which are fermionic in nature and should therefore obey the
commutation rules, $[\Gamma^\alpha,\Gamma^{\bar{\beta}}]=2I^{\alpha\bar{\beta}}$, 
instead of the usual Clifford ones. We show how to do this using a particular 
representation that entrains auxiliary anticommuting variables $\theta$, in Appendix A.5.
The net conclusion of that appendix is that we may discard the effect of the new set
of  Grassmannian gamma-matrices $\Gamma^\alpha$, provided that $\Psi$ is
a singlet $\Theta$ of the new representation and carries that factor $\Theta$.

Thus we take write the anti-selfdual fermionic superfield and its adjoint in flat space to be
\begin{equation}
\Psi=(\bar{\zeta}\psi +\psi^c\zeta)(1-\bar{\zeta}\zeta)\Theta/2,\quad
\bar{\Psi}=(-\bar{\psi}\zeta+\bar{\zeta}\overline{\psi^c})(1-\bar{\zeta}\zeta)\bar{\Theta}/2,
\end{equation}
where $\bar{\zeta}\psi\equiv \zeta^{\bar{1}}\psi^1+ \zeta^{\bar{2}}\psi^2$ and $\psi^c$
is the charge conjugate of $\psi$.
In that way we check that after integrating over auxiliary $\theta$, the mass and 
kinetic terms arise in the usual fashion:
\begin{equation}
\int\!\! d^2\zeta d^2\bar{\zeta}\,\bar{\Psi}\Psi = \!\!\int \!\!d^2\zeta d^2\bar{\zeta}\,(\bar{\zeta}\zeta)
      (1\!-\!2\bar{\zeta}\zeta)[\bar{\psi}\psi+\overline{\psi^c}\psi^c]/4= -2\bar{\psi}\psi,
\end{equation}
\begin{equation}
\int \!\!d^2\zeta d^2\bar{\zeta}\,\bar{\Psi}i\gamma.\partial\Psi=\!\!\int\!\! d^2\zeta d^2\bar{\zeta}\,
  (\bar{\zeta}\zeta) (1-2\bar{\zeta}\zeta)[\bar{\psi}i\gamma.\partial\psi+
  \overline{\psi^c}i\gamma.\partial\psi^c]/4=-2\bar{\psi}i\gamma.\partial\psi.
\end{equation}
We now proceed to curve the space through the frame vectors $E$, ignoring property 
curvature coefficients except for $c_3=c$ and  focussing on the gauge field which lives 
in the property-spacetime sector, viz.
\[
{E_A}^M\partial_M=[1+c\bar{\zeta}\zeta/2]({e_a}^m\partial_m+ie(W_a\zeta)^\mu\partial_\mu
   -ie(\bar{\zeta}W_a)^{\bar{\mu}}\partial_{\bar{\mu}}+`0').
\]
The gauge field produces a derivative over the property, but also introduces a compensating 
property factor and the net result is that
\begin{equation}
-\!\!\int\!\!d^2\zeta d^2\bar{\zeta}(s\det{E})\bar{\Psi}i\Gamma^A\!{E_A}^M\partial_M\Psi\!=\!
\frac{8\sqrt{g..}}{l^4}(1\!-\!\frac{c}{4})\!\left[\bar{\psi}\gamma.(i\partial\!-\!eW\!)\psi\!+\!
  \overline{\psi^c}\gamma.(i\partial\!+\!eW\!)\psi^c\!\right].
\end{equation}
This is exactly what one would have anticipated and it confirms that the frame vector and
resulting metric have precisely the right forms not only for scalar sources but for 
spinor ones too.

\section{Conclusions}
In this article we have proved, with two properties and a metric that possesses the
correct structure in the property-spacetime sectors (for incorporating gauge transformations),
that the Yang-Mills Lagrangian automatically follows upon integrating over the property
sector. We have also shown that the interaction of the gauge field with scalar and spinor
fields assumes the correct form. The inclusion of spacetime curvature merely makes the
formalism $x$-coordinate independent and conform to standard GR. 

What we have not done in this paper
is treated the case where the spinor left and right chiralities behave differently, 
as in electroweak theory. This is left to for a subsequent paper because it brings in 
new concepts; it is technically quite difficult because one is dealing with four 
properties actually  -- two for the lepton and a further two for the neutrino -- and 
has the added complication of large numbers of property invariants.
Until we are able to systematise and simplify the number of $c_i$ coefficients we must 
leave this subject in abeyance for now. However one point needs making before
departing: it is that one has to pick the $ce^2$ for each gauge interaction to have
a uniform value in order to guarantee gravitational universality; this becomes
a significant constraint when the property symmetry group is a direct product of subgroups.
 
\section{Acknowledgements}  
Regular meetings with Dr Peter Jarvis have  helped to illuminate a lot of issues.
 
%\clearpage
\appendix
\section{Frame vectors, Inverse metric, Christoffel symbols and Ricci tensor components}
\subsection{Frame vectors}
We will define the following to include property curvature in our frame vectors: $T = \big(1+c_3 \bar\zeta\zeta + c_4 (\bar\zeta\zeta)^2\big)$ and 
$ S = \big( 1 + [c_1 - c_3] \bar\zeta\zeta + [c_2 - c_4 -c_3 (c_1 - c_3)](\bar\zeta\zeta)^2\big) $. 
Note that $S T = (1+c_1 \bar\zeta\zeta + c_2 (\bar\zeta\zeta)^2)$.\\
 We adopt upper-triangular frame vectors of the form:
\begin{equation}
\mathcal{E}_M{}^{A} =
\sqrt{T} \left( \begin{array}{ccc}
\sqrt{S} e_m{}^a & - i e W_m{}^{ \alpha \bar\nu} \zeta^{\nu} &  i e \zeta^{\bar\nu} W_m{}^{ \nu \bar\alpha} \\
0 & \delta_\mu{}^\alpha &  0\\
0 & 0 &\delta_{\bar\mu}{}^{\bar\alpha} \\
\end{array}
\right)
\end{equation}
with inverse vielbein,
\begin{equation}
E_A{}^{M} =
\frac{1}{\sqrt{T}}
\left(
\begin{array}{ccc}
(\sqrt{S})^{-1} e_a{}^m &  i (\sqrt{S})^{-1} e  W_a{}^{ \mu \bar\nu} \zeta^{\nu} & - i (\sqrt{S})^{-1} e \zeta^{\bar\nu} W_a{}^{ \nu \bar\mu} \\
0 & \delta_\alpha{}^\mu &  0\\
0 & 0 &\delta_{\bar\alpha}{}^{\bar\mu} 
\end{array}
\right).
\end{equation}
These of course satisfy
\begin{equation}
\mathcal{\cal E}_{M}{}^{B} E_{B}{}^{N} = \delta_{M}{}^{N}.
\end{equation}
The metric is produced via $G_{MN} = (-1)^{[A] [N]} {\cal E}_M{}^{A} {\cal E}_N{}^B \mathcal{I}_{B A}.$
\subsection{Inverse metric}
 The inverse metric is found in a similar fashion to the metric through
\begin{equation}
G^{MN} = (-1)^{[B][M]} \mathcal{I}^{BA} E_A{}^{M} E_B{}^N;
\end{equation}
this results in the following inverse metric:
\begin{equation}
G^{MN} = 
\frac{1}{ST}
\left(
\begin{array}{ccc}
 g^{mn} & i e (W^{m}\zeta)^{\nu} & - i e (\bar\zeta W^{m})^{\bar\nu} \\ 
i e (W^{n}\zeta)^{\mu} & - e{}^2 (W^{k}\zeta){}^{\mu} (W_{k}\zeta){}^{\nu} 
&\frac{2}{l^2} S \delta{}^{\mu \bar{\nu}}\!\!-\! e{}^2 (\bar\zeta W^{k}){}^{\bar{\nu}}(W_{k}\zeta){}^{\mu}  \\ 
 - ie(\bar\zeta W^{n})^{\bar\mu}&-\frac{2}{l^2} S \delta{}^{\nu \bar{\mu}}\!\!+\!e{}^2 (\bar\zeta W^{k}){}^{\bar{\mu}} (W_{k}\zeta){}^{\nu} & - e{}^2 (\bar\zeta W^{k}){}^{\bar{\mu}} (\bar\zeta W_{k}){}^{\bar{\nu}}
\end{array}
\right).
\end{equation}
Expanding out $T$ and $S$ give the following list of components of the inverse metric:
\begin{align*}
G^{mn} =& g^{mn} \big[1 - c_1  \bar\zeta \zeta  + (c_1^2 - c_2) (\bar\zeta \zeta)^2 \big],  \nonumber\\
G^{m\nu} =&  i e (W^{m}\zeta)^{\nu}\big[1 - c_1  \bar\zeta \zeta\big] , \nonumber \\
G^{m\bar\nu} =&  - i e (\bar\zeta W^{m})^{\bar\nu}\big[1 - c_1  \bar\zeta \zeta\big], \nonumber \\
G^{\mu\nu} =& - e{}^2 (W^{k}\zeta){}^{\mu} (W_{k}\zeta){}^{\nu}\big[1 - c_1  \bar\zeta \zeta\big], \nonumber \\
G^{\bar\mu\bar\nu} =& - e{}^2 (\bar\zeta W^{k}){}^{\bar{\mu}} (\bar\zeta W_{k}){}^{\bar{\nu}}\big[1 - c_1  \bar\zeta \zeta\big],\nonumber \\
G^{\mu \bar\nu} =&  \frac{2}{l^2} \delta{}^{\mu \bar{\nu}}  
\big[1 - c_3  \bar\zeta \zeta  + (c_3^2 - c_4) (\bar\zeta \zeta)^2 \big]
- e{}^2 (\bar\zeta W^{k}){}^{\bar{\nu}}(W_{k}\zeta){}^{\mu} \big[1 - c_1  \bar\zeta \zeta\big] .
\end{align*}

Note that the metric and its inverse are still graded symmetric, 
$G_{MN} = (-1)^{[M][N]} G_{NM}$ and $G^{MN} = (-1)^{[M][N]} G^{NM}$. The $G^{\mu\nu}$ 
and $G^{\bar\mu\bar\nu}$ sectors of the inverse metric are no longer one by one like 
they were in the abelian case, so they can be non zero without breaking the symmetry.

\subsection{Christoffel symbols}
Using (\ref{gammaformula}) and the metric from (\ref{2coordmetric}) we calculated the 
following list of Christoffel symbols. All of the algebra was assisted by use of Mathematica.
\\
\\
\(
\Gamma_{mn}{}^{l} = \Gamma^{[g]}{}_{mn}{}^{l}-\frac{1}{4} e{}^2 l^2  g{}^{l}{}^{k}  \big(1+ (c_3 - c_1) \bar\zeta \zeta \big)  \bar\zeta [W_{m}W_{n}{}_{,k}-W_{m}W_{k}{}_{,n}+ W_{m}{}_{,k}W_{n}-  W_{k}{}_{,m}W_{n}+i e W_{m}W_{n}W_{k}- i e W_{k}W_{m}W_{n} + (m \leftrightarrow n) ] \zeta,
\)
\\
\\
\(
\Gamma_{mn}{}^{\lambda} =  i e \Gamma^{[g]}{}_{m n}{}^{k} (W_{k}\zeta){}^{\lambda} - \frac{1}{l^2} g{}_{m}{}_{n} \zeta^{\lambda} [c_1  + (2  c_2 - c_1 c_3 ) \bar\zeta \zeta ] 
-\frac{1}{4} e{}^2  l^2 ( W^{k} \zeta){}^{\lambda} \bar\zeta \big[\\ e W_{k}W_{m}W_{n} - e W_{m}W_{n}W_{k} + i W_{m}W_{n}{}_{,k}- i W_{m}W_{k}{}_{,n}+ i W_{m}{}_{,k}W_{n}- i W_{k}{}_{,m}W_{n}  +(m \leftrightarrow n)\big]\zeta 
-\frac{1}{2} c_3 e{}^2 \bar\zeta (W_{m}W_{n}+W_{n}W_{m})\zeta  \zeta^{\lambda}  
-\frac{1}{2} e \big[ (e W_{m}W_{n}+ i  W_{m}{}_{,n})\zeta  +(m \leftrightarrow n) \big]{}^{\lambda}{},
\)
\\
\\
\(
\Gamma_{mn}{}^{\bar\lambda} = - i e \Gamma^{[g]}{}_{m n}{}^{k} (\bar\zeta W_{k}){}^{\bar{\lambda}}
-\frac{1}{l^2} g{}_{m}{}_{n} \zeta^{\bar{\lambda}}[  c_1 + (2 c_2- c_1 c_3) \bar\zeta\zeta]
+\frac{1}{4} e{}^2 l^2   (\bar\zeta W^{k}){}^{\bar{\lambda}} \bar\zeta \big[\\ (e W_{k}W_{m}W_{n} - e W_{m}W_{n}W_{k}+ i W_{m}W_{n}{}_{,k}- i W_{m}W_{k}{}_{,n}+ i W_{m}{}_{,k}W_{n}- i W_{k}{}_{,m}W_{n} +(m \leftrightarrow n) \big]\zeta 
- \frac{1}{2} c_3 e{}^2 \bar\zeta (W_{m}W_{n}+W_{n}W_{m})\zeta  \zeta^{\bar{\lambda}} -\frac{1}{2} e \big[ \bar\zeta ( e W_{m}W_{n} - i W_{m}{}_{,n}) + (m \leftrightarrow n)   \big]{}^{\bar{\lambda}} {},
\)
\\
\\
\(
\Gamma_{m\nu}{}^{l} = 
-\frac{1}{2} [ c_1+(2 c_2 -c_1{}^{2}) (\bar\zeta\zeta) ]\delta{}_{m}{}^{l} \zeta^{\bar{\nu}}\\
-\frac{1}{4} e l^2 g{}^{l}{}^{k} \bar\zeta [e W_{m}W_{k} - i W_{m}{}_{,k} - (m \leftrightarrow k) ]{}^{\bar{\nu}}[1+ (c_3- c_1)(\bar\zeta \zeta)],
\)
\\
\\
\(
\Gamma_{m \bar\nu}{}^{l} = 
\frac{1}{2} \delta{}_{m}{}^{l}\zeta^{\nu}  [c_1 +(2 c_2- c_1^2) (\bar\zeta \zeta)  ]\\
-\frac{1}{4} e l^2 g{}^{l}{}^{k}  \bigg( e [(W_{m}W_{k}-W_{k}W_{m})\zeta]^{\nu}-i [(W_{m}{}_{,k}-W_{k}{}_{,m})\zeta] {}^{\nu}\bigg)[1 -  (c_1-c_3)(\bar\zeta \zeta)],
\)
\\
\\
\(
\Gamma_{m \nu}{}^{\lambda} = 
\frac{i}{2} e [\bar\zeta W_{m}]{}^{\bar \nu}\zeta^{\lambda} c_3 (1-c_3 (\bar\zeta \zeta) )
+\frac{i}{2} e  \zeta^{\bar{\nu}}[W_{m}\zeta] {}^{\lambda} \bigg(c_3-c_1- (2 c_2+c_3{}^2 - c_1{}^2)(\bar\zeta \zeta)\bigg)\\
- i e  (W_{m}){}^{\lambda}{}^{\bar{\nu}} \bigg(1-c_4 (\bar\zeta \zeta)^2 \bigg)
+ \frac{i}{4} e{}^2 l^2  [W^{k}\zeta] {}^{\lambda} \bigg( e [\bar\zeta (W_{m}W_{k}-W_{k}W_{m})]{}^{\bar \nu} \\
- i e  [\bar\zeta (W_{m}{}_{,k}-W_{k}{}_{,m})]{}^{\bar \nu}\bigg) \bigg(1 +(c_3-c_1)(\bar\zeta \zeta)   \bigg),
\)
\\
\\
\(
\Gamma_{m \bar\nu}{}^{\bar\lambda} = 
-\frac{i}{2} e \zeta^{\bar{\lambda}}[W_{m}\zeta] {}^{\nu} c_3 [1- c_3 (\bar\zeta \zeta) ]
-\frac{i}{2} e [\bar\zeta W_{m}]{}^{\bar \lambda}\zeta^{\nu} \bigg(c_3-c_1-(2 c_2 + c_3{}^2 - c_1{}^2) (\bar\zeta \zeta)\bigg)\\
+i e (W_{m}){}^{\nu}{}^{\bar{\lambda}} [1-  c_4 (\bar\zeta \zeta)^2]
-\frac{i}{4} e{}^2 l^2  [\bar\zeta W^{k}]{}^{\bar \lambda} \bigg( e [(W_{m}W_{k}-W_{k}W_{m})\zeta] {}^{\nu} \\
- i [(W_{m}{}_{,k}-W_{k}{}_{,m})\zeta] {}^{\nu}\bigg) [1+(c_3 - c_1) (\bar\zeta \zeta) ],
\)
\\
\\
\(
\Gamma_{m \nu}{}^{\bar\lambda} = 
-\frac{1}{4} e{}^2 l^2  [\bar\zeta W^{k}]{}^{\bar \lambda} \bigg( [\bar\zeta (W_{m}{}_{,k}-W_{k}{}_{,m})]{}^{\bar \nu}+i e [\bar\zeta (W_{m}W_{k}-W_{k}W_{m})]{}^{\bar \nu}\bigg)\\
+\frac{1}{2} i e \bigg(\zeta^{\bar{\nu}}[\bar\zeta W_{m}]{}^{\bar \lambda} (c_1-c_3)-\zeta^{\bar{\lambda}}[\bar\zeta W_{m}]{}^{\bar \nu} c_3\bigg),
\)
\\
\\
\(
\Gamma_{m \bar\nu}{}^{\lambda} = 
\frac{1}{4} e{}^2 l^2  [W^{k}\zeta] {}^{\lambda} \bigg([(W_{m}{}_{,k}-W_{k}{}_{,m})\zeta] {}^{\nu}+i e [(W_{m}W_{k}-W_{k}W_{m})\zeta] {}^{\nu}\bigg)\\
+\frac{1}{2}  i e \bigg(\zeta^{\nu}[W_{m}\zeta] {}^{\lambda} (c_1-c_3)-\zeta^{\lambda}[W_{m}\zeta] {}^{\nu} c_3\bigg),
\)
\\
\\
\(
\Gamma_{\mu \nu}{}^{\lambda} = 
-\frac{1}{2}\zeta^{\bar{\nu}} (c_3{}^2-2 c_4)(\bar\zeta \zeta) \delta^{\lambda\bar \mu} 
-\frac{1}{2}\delta^{\lambda\bar \nu}  \zeta^{\bar{\mu}} [c_3 - (c_3{}^2-2 c_4)(\bar\zeta \zeta) ]
+\frac{1}{2} c_3 \delta^{\lambda\bar \mu}  \zeta^{\bar{\nu}}{},
\)
\\
\\
\(
\Gamma_{\bar\mu \bar\nu}{}^{\bar\lambda} = 
\frac{1}{2} \zeta^{\nu}  \delta^{\mu\bar \lambda}  (c_3{}^2-2 c_4) (\bar\zeta \zeta)
+\frac{1}{2} \delta^{\nu\bar \lambda} \zeta^{\mu}   [c_3 - (c_3{}^2-2 c_4) (\bar\zeta \zeta) ]
-\frac{1}{2}c_3 \delta^{\mu\bar \lambda}  \zeta^{\nu}{},
\)
\\
\\
\(
\Gamma_{\mu \bar\nu}{}^{l} = 
-\frac{i}{2} e l^2 c_4 \bigg(\zeta^{\bar{\mu}}[W^{l}\zeta] {}^{\nu}-[\bar\zeta W^{l}]{}^{\bar \mu}\zeta^{\nu}\bigg) (\bar\zeta \zeta) ,
\)
\\
\\
\(
\Gamma_{\mu \bar\nu}{}^{\lambda} = 
\frac{1}{2} (c_3{}^2-2 c_4)(\bar\zeta \zeta)    \zeta^{\nu}\delta^{\lambda\bar \mu} 
+\frac{1}{2}  (c_3{}^2-2 c_4) (\bar\zeta \zeta)\zeta^{\lambda} \delta^{\nu\bar \mu} 
-\frac{1}{2}c_3 (\delta^{\nu\bar \mu}  \zeta^{\lambda}+\delta^{\lambda\bar \mu}  \zeta^{\nu}){},
\)
\\
\\
\(
\Gamma_{\mu \bar\nu}{}^{\bar\lambda} = 
 \frac{1}{2} (c_3{}^2-2 c_4)(\bar\zeta \zeta) \zeta^{\bar{\mu}} \delta^{\nu\bar \lambda} 
+\frac{1}{2}  (c_3{}^2-2 c_4) (\bar\zeta \zeta) \zeta^{\bar{\lambda}} \delta^{\nu\bar \mu}
 -\frac{1}{2} c_3 (\delta^{\nu\bar \mu}  \zeta^{\bar{\lambda}}+\delta^{\nu\bar \lambda}  \zeta^{\bar{\mu}}),
\)
\\
\\
\(
\Gamma_{\mu \nu}{}^{l} = \Gamma_{\mu \nu}{}^{\bar\lambda} = \Gamma_{\bar\mu \bar\nu}{}^{l} = \Gamma_{\bar\mu \bar\nu}{}^{\lambda} = 0.
\)

\subsection{Ricci tensor components}
Here we list the components of the Ricci tensor $R^{MN}$. To simplify these expressions we have ignored space-time curvature and set $c_1 = c_2 = c_4 = 0$ with $c_3 = c$. (Readers who wish to see the full set, including all other $c_i$ should contact P.D.Stack directly.) 
Note that $F_{mn} = W_{n,m} - W_{m,n}$ and $\mathcal{F}_{mn} = F_{mn} - i e [ W_m, W_n]$.
\\
\\
\\
\(
R^{mn}=
-\frac{1}{4}e^2 l^2 [\bar\zeta  (\mathcal{F}{}^{m}{}_{k}\mathcal{F}{}^{n}{}^{k} ) \zeta]-c^2 (\bar\zeta \zeta) 
[\bar\zeta (W{}^{m}W{}^{n}) \zeta]\\
+\frac{1}{4} ce^3l^2( [\bar\zeta W{}_{k}\zeta] - (\bar\zeta \zeta) \text{Tr}(W{}_{k}) 
[\bar\zeta (eW{}^{k}W{}^{m}W{}^{n}+eW{}^{m}W{}^{n}W{}^{k}-2eW{}^{m}W{}^{k}W{}^{n}
-i (F{}^{m}{}^{k}W{}^{n} -W{}^{m}F{}^{n}{}^{k})) \zeta]
-e^2c^2 ( [\bar\zeta W{}^{m}\zeta] - (\bar\zeta \zeta) \text{Tr}(W{}^{m}))
 [\bar\zeta W{}^{n}\zeta]\\
-\frac{i}{4} ce^2 l^2 (\bar\zeta \zeta) 
[\bar\zeta (eF{}^{m}{}_{k}W{}^{k}W{}^{n}-eW{}^{m}W{}_{k}F{}^{n}{}^{k}
-i (F{}^{m}{}_{k}F^{n}{}^{k}-e^2W{}_{k}W{}^{m}W{}^{n}W{}^{k}+e^2W{}^{m}W{}_{k}W{}^{k}W{}^{n}))\zeta] 
+ (m \leftrightarrow n),
\)
\\
\\
\(
R^{\mu \nu} = -\frac{9c^2}{l^4} \zeta^{\mu}\zeta^{\nu}\\
\!\!+\!\!\frac{1}{4}e^2[W{}_{k}{}_{,m}\zeta] {}^{\mu}[ (F{}^{k}{}^{m}\!\!-\!2ie[W^{k},W^{m}] )\zeta] {}^{\nu}
+\!\frac{1}{4} e^4[W{}_{k}W{}_{m}\zeta] {}^{\mu}[[W{}^{k},W{}^{m}] \zeta] {}^{\nu}\\
-\!\!\frac{i}{2} e^2[W{}_{k}\zeta] {}^{\mu}[(\!2eW{}^{k}{}_{,m}W{}^{\!m}\!\!\!-\!\!2eW{}_{\!m}W{}^{k}{}^{,m}
\!\!-\!eW{}_{m}{}^{,k}W{}^{m}\!\!-\!\!eW{}_{\!m}{}^{,m}W{}^{k}\!\!+\!\!eW{}_{\!m}W{}^{m}{}^{,k}
\!+\!eW{}^{k}W{}_{\!\!m}{}^{,m} \\
- 2 ie^2 W{}_{\!m}W{}^{k}W{}^{m} +ieW{}_{\!m}{}^{,k}{}^{,m}- i W{}^{k}{}_{,m}{}^{,m}
+ ie^2W{}_{\!m}W{}^{m}W{}^{k} + i e^2W{}^{k}W{}_{\!m}W{}^{m}))\zeta] {}^{\nu}\\
- (\mu \leftrightarrow \nu),
\)
\\
\\
\(
R^{\bar\mu \bar\nu}=
-\frac{9c^2}{l^4} \zeta^{\bar{\mu}}\zeta^{\bar{\nu}}\\
+\frac{1}{4}e^2 [\bar\zeta W{}_{k}{}_{\!,m}]{}^{\bar \mu}[\bar\zeta  (F{}^{k}{}^{m}\!-\!2ie[W^k,W^m])]{}^{\bar \nu}
+\frac{1}{4}e^4 [\bar\zeta W{}_{k}W{}_{m}]{}^{\bar \mu}[\bar\zeta [W{}^{k},W{}^{m}]]{}^{\bar \nu}\\
-\!\!\frac{i}{2}e^2[\bar\zeta W{}_{k}]{}^{\bar \mu}[\bar\zeta (\!2eW{}^{k}{}_{\!,m}W{}^{m}\!\!-\!\!
 2eW{}_{\!m}W{}^{\!k}{}^{\!,m}\!\!-\!\!eW{}_{\!m}{}^{\!,k}W{}^{\!m}\!-\!eW_{\!m}{}^{\!,m}W{}^{k}\!+
 \!eW{}_{\!m}W{}^{\!m}{}^{\!,k}\!+\!eW{}^{\!k}W{}_{\!m}{}^{\!,m}\\
 -\!2ie^2 W{}_{\!m}W{}^{k}W{}^{\!m}\!+\!ieW{}_{m}{}_{\!,k}{}^{,m}\!-\!i W{}^{k}{}_{,m}{}^{,m}\!+
 \!ie^2 W{}_{\!m}W{}^{\!m}W{}^{k} \!+\!ie^2W{}^{k}W{}_{m}W{}^{m})]{}^{\bar \nu}\\
- (\bar\mu \leftrightarrow \bar\nu),
\)
\\
\\
\(
R^{m \nu} =
\frac{1}{2} ce^2 (\bar\zeta \zeta) \text{Tr}(W{}_{k}) [ \mathcal{F}{}^{m}{}^{k}\zeta] {}^{\nu}
-\frac{1}{2} c e^2(\bar\zeta \zeta) [\mathcal{F}^{m}{}_{k}  W{}^{k}\zeta] {}^{\nu}\\
-i c^2 e\frac{1}{l^2} (\bar\zeta \zeta) [W{}^{m}\zeta] {}^{\nu}
-i c^2e\frac{1}{l^2} [\bar\zeta W{}^{m} \zeta] \zeta^{\nu}
+i c^2e \frac{1}{l^2} (\bar\zeta \zeta) \text{Tr}(W{}^{m}) \zeta^{\nu}\\
-\frac{1}{2} ce^2   [\bar\zeta W{}_{k}\zeta][  \mathcal{F}{}^{m}{}^{k}\zeta]{}^{\nu}
-\frac{i}{4} l^2e^3  [\bar\zeta  (\mathcal{F}{}^{k}{}_{l}\mathcal{F}{}^{m}{}^{l}+
\mathcal{F}{}^{m}{}_{l}\mathcal{F}{}^{k}{}^{l}) \zeta][W{}_{k} \zeta] {}^{\nu}\\
-\frac{1}{2}e [(2eW{}^{m}{}_{,k}W{}^{k}-2eW{}_{k}W{}^{m}{}^{,k}-eW{}_{\!k}{}^{,k}W{}^{m}
-eW{}_{k}{}^{,m}W{}^{k}+eW{}_{k}W{}^{k}{}^{,m}+eW{}^{m}W{}_{k}{}^{,k}\\
 -2 ie^2W{}_{k}W{}^{m}W{}^{k} +i W{}_{k}{}^{,m}{}^{,k}-
 i W{}^{m}{}_{,k}{}^{,k}+ie^2W{}_{k}W{}^{k}W{}^{m}+ie^2W{}^{m}W{}_{k}W{}^{k})\zeta] {}^{\nu},
\)
\\
\\
\(
R^{m \bar\nu} = 
\frac{1}{2} ce^2 (\bar\zeta \zeta)\text{Tr}(W{}_{k}) [\bar\zeta  \mathcal{F}{}^{m}{}^{k}]{}^{\bar \nu} 
-\frac{1}{2} c e^2(\bar\zeta \zeta) [\bar\zeta  W{}_{k} \mathcal{F}{}^{m}{}^{k}]{}^{\bar \nu}\\
+i c^2e \frac{1}{l^2}(\bar\zeta \zeta) [\bar\zeta W{}^{m}]{}^{\bar \nu}
+i c^2e \frac{1}{l^2}[\bar\zeta W{}^{m}\zeta] \zeta^{\bar{\nu}} 
-i c^2e \frac{1}{l^2} (\bar\zeta \zeta) \text{Tr}(W{}^{m})\zeta^{\bar{\nu}}\\
-\frac{1}{2} ce^2[\bar\zeta W{}^{k} \zeta] [\bar\zeta  \mathcal{F}{}^{m}{}_{k}]{}^{\bar \nu}
+\frac{i}{4} l^2e^3 [\bar\zeta  (\mathcal{F}{}^{k}{}_{l}\mathcal{F}{}^{m}{}^{l}+
\mathcal{F}{}^{m}{}_{l}\mathcal{F}{}^{k}{}^{l}) \zeta] [\bar\zeta W{}_{k}]{}^{\bar \nu}\\
+\frac{1}{2}e[\bar\zeta (2eW{}^{m}{}_{,k}W{}^{k}- 2eW{}_{k}W{}^{m}{}^{,k}
-eW{}_{k}{}^{,k}W{}^{m}-eW{}_{k}{}^{,m}W{}^{k}+eW{}_{k}W{}^{k}{}^{,m}+eW{}^{m}W{}_{k}{}^{,k}\\
- 2 ie^2W{}_{k}W{}^{m}W{}^{k} +i W{}_{k}{}^{,m}{}^{,k} -i W{}^{m}{}_{,k}{}^{,k} 
+i e^2W{}_{k}W{}^{k}W{}^{m} +ie^2W{}^{m}W{}_{k}W{}^{k})]{}^{\bar \nu},
\)
\\
\\
\(
R^{\mu \bar\nu} = 
\frac{1}{l^4} [20 c-44 c^2 (\bar\zeta \zeta)+44 c^3 (\bar\zeta \zeta)^2] \delta^{\mu\bar \nu} 
+\frac{1}{l^4} [18 c^2-48 c^3 (\bar\zeta \zeta)] \zeta^{\bar{\nu}}\zeta^{\mu}\\
+\frac{1}{4}e^4 l^2 [\bar\zeta W{}_{k} ]{}^{\bar \nu}[W{}_{m} \zeta] {}^{\mu} 
\bar\zeta  (\mathcal{F}{}^{k}{}_{l}\mathcal{F}{}^{m}{}^{l}+\mathcal{F}{}^{m}{}_{l}\mathcal{F}{}^{k}{}^{l}) \zeta]\\
+\frac{1}{2}e^2 [\bar\zeta W{}_{k}{}_{,m}]{}^{\bar \nu}[ \mathcal{F}{}^{k}{}^{m}\zeta] {}^{\mu}
+2 c^2e^2 \frac{1}{l^2} (\bar\zeta \zeta) [\bar\zeta W{}_{k}]{}^{\bar \nu}[W{}^{k}\zeta] {}^{\mu}\\
+c^2e^2 \frac{1}{l^2}( [\bar\zeta W{}^{k} ]{}^{\bar \nu}\zeta^{\mu}
+\zeta^{\bar{\nu}}[W{}_{k} \zeta] {}^{\mu})( [\bar\zeta W{}_{k} \zeta ] -  (\bar\zeta \zeta)  \text{Tr}(W{}_{k} )) \\
+\frac{i}{2} ce^3( [\bar\zeta W{}_{m}]{}^{\bar \nu}[ \mathcal{F}{}^{k}{}^{m}\zeta] {}^{\mu} - 
[\bar\zeta  \mathcal{F}{}^{k}{}^{m}]{}^{\bar \nu}[W{}_{m}\zeta] {}^{\mu})( [\bar\zeta W{}_{k}\zeta] - (\bar\zeta \zeta) \text{Tr}(W{}_{k} ))\\
+ \frac{i}{2}ce^3 (\bar\zeta \zeta) [\bar\zeta  W{}_{m}\mathcal{F}{}_{k}{}^{m} ]{}^{\bar \nu}[W{}^{k}\zeta] {}^{\mu}
+\frac{i}{2} ce^3(\bar\zeta \zeta)[ \mathcal{F}{}^{k}{}_{m} W{}^{m}\zeta] {}^{\mu} [\bar\zeta W{}_{k}]{}^{\bar \nu}\\
-\frac{i}{2}e^3 [\bar\zeta [W{}_{k},W{}_{m}] ]{}^{\bar \nu}[W{}^{k}{}^{,m} \zeta] {}^{\mu}
+\frac{1}{2}e^4 [\bar\zeta W{}_{k}W{}_{m}]{}^{\bar \nu}[[W{}^{k},W{}^{m}]\zeta] {}^{\mu}\\
-\frac{i}{2}e^2 [\bar\zeta W{}_{k} ]{}^{\bar \nu}[(\!2e W{}^{k}{}_{\!,m}W{}^{\!m}\!-\!2e W{}_{m}W{}^{k}{}^{,m} 
-eW{}_{\!m}{}^{,k}W{}^{\!m}\!-\!eW{}_{\!m}{}^{\!,m}W{}^{\!k}\!+\!eW{}_{m}W{}^{m}{}^{,k}\!+\!eW{}^{k}W{}_{m}{}^{,m}
\!-\!2 ie^2W{}_{\!m}W{}^{k}W{}^{\!m} +i{W{}_{\!m}}^{,k,m}\!-\!i W{}^{k}{}_{,m}{}^{,m}\!+
\!ie^2W{}_{\!m}W{}^{\!m}W{}^{k}\!+\!i e^2W{}^{k}W{}_{\!m}W{}^{\!m})\zeta] {}^{\mu}\\
+\frac{i}{2}e^2[W{}^{k}\zeta] {}^{\mu} [\bar\zeta (2eW{}_{k}{}_{,m}W{}^{m}-2e W{}_{m}W{}_{k}{}^{,m}
-eW{}_{\!m}{}_{\!,k}W{}^{\!m}\!-\!eW{}_{\!m}{}^{\!,m}W{}_{k}\!+\!eW{}_{\!k}W{}_{m}{}^{,m}\\
\!+eW{}_{\!m}W{}^{m}{}_{,k}\!-\!\!2 ie^2 W{}_{\!m}W{}_{\!k}W{}^{\!m}\!-\!iW{}_{\!k}{}_{\!,m}{}^{,m}\!\!+
\!iW{}_{\!m}{}_{,k}{}^{,m} \!+ie^2W{}_{\!k}W{}_{\!m}W{}^{m}\!+ie^2W{}_{\!m}W{}^{\!m}W{}_{k})]{}^{\bar \nu}.
\)

\subsection{Square Rooting the Grassmann Metric}
Let $\Gamma^AP_A$ stand for the Dirac operator in flat space, where $P_A\equiv i\partial/\partial X^A$.
The Dirac operator has to be bosonic like mass. 
Now $\Gamma^a\equiv \gamma^a$ and $P_a$ which act in space-time 
are both bosonic, but $P_\alpha$ which serves as the Grassmann derivative is fermionic, 
so we require $\Gamma^\alpha$ to be fermionic too.
Since we also demand that $(\Gamma^AP_A)^2 = \eta^{AB}P_BP_A$, we deduce that in the
property sector,
\[
[\Gamma^\alpha,\Gamma^\beta]=[\Gamma^{\bar{\alpha}},\Gamma^{\bar{\beta}}] =0, \quad
[\Gamma^\alpha,\Gamma^{\bar{\beta}}]= 2\eta^{\alpha\bar{\beta}}=2{\delta_\beta}^\alpha.
\]
The first two commutators can be guaranteed by doubling the space and writing
$\Gamma^\alpha=\sigma_+{\cal O}^\alpha,\, \Gamma^{\bar{\alpha}}=\sigma_-{\cal O}^{\bar{\alpha}}$
since $\sigma_+^2=\sigma_-^2=0$. This then leaves
\[
[\Gamma^\alpha,\Gamma^{\bar{\beta}}] =\sigma_3\{{\cal O}^\alpha,{\cal O}^{\bar{\beta}}\}/2 +
[{\cal O}^\alpha,{\cal O}^{\bar{\beta}}]/2, 
\]
which must somehow be reduced to $2{\delta_\beta}^\alpha$.

Now these $\Gamma^\alpha$ or ${\cal O}^\alpha$ must carry the same number of attribute labels
as $P_\alpha$ -- 2 for U(2) -- and be somehow distinguished from the original property coordinates
$\zeta^\alpha$. One way to ensure this is to observe that if we introduce auxiliary
a-scalars, $\Gamma^\alpha\equiv\theta^\alpha$ and 
$\Gamma^{\bar{\alpha}}\equiv \partial/\partial \theta^\alpha$, with
\[
\{\theta^\alpha,\partial/\partial\theta^\beta\}={\delta_\beta}^\alpha,\quad
[\theta^\alpha,\partial/\partial\theta^\beta]={\delta_\beta}^\alpha-2(\partial/\partial\theta^\beta)\theta^\alpha
\]
and make these act on {\em singlets}\, $\Theta=\theta^1\theta^2\ldots\theta^N$, we attain our goal. All told 
then $[\Gamma^\alpha,\Gamma^{\bar{\beta}}]=(1+\sigma_3){\delta_\beta}^\alpha$, when
projected on to the singlet $\Theta$. [There may be other ways of `square-rooting' the Grassmann metric
but our procedure more or less does the job.] We make use of this representation of $\Gamma^\alpha$
and $\Gamma^{\bar{\alpha}}$ in section 4.2. The net result is that if we multiply $\Psi$ by $\Theta$ 
and integrate over auxiliary $\theta$-space the part of the Dirac operator which involves
$\Gamma^\alpha$ vanishes.


\begin{thebibliography}{99}

\bibitem{RDPJRW} R.~Delbourgo, P.D.~Jarvis and R.~Warner, J. Math. Phys. 
{\bf 34}, 3616 (1993)
\bibitem{RDRZ} R.~Delbourgo and R.B.~Zhang, Phys. Rev. {\bf D38}:2490 (1988).
\bibitem{RD2} R.~Delbourgo, J. Phys. {\bf A39}, 5175 (2006); {\em ibid}, 14735 (2006)
\bibitem{RD3} R.~Delbourgo, Int. J. Mod. Phys. {\bf 22A}, 29 (2007).
\bibitem{RD4} R.~Delbourgo, arXiv:1202.4216 (2012).
\bibitem{empaper} R.~Delbourgo and P.~Stack, Int. J. Mod. Phys. {\bf 29A}, 50023 (2014).


 \end{thebibliography}
\end{document}